\documentclass[12pt,twoside]{article} 

\usepackage{geometry}

\date{} 

\RequirePackage{graphicx}
\usepackage{amssymb}
\usepackage{hyperref}
\usepackage{fancyhdr}
\usepackage{tabulary}
\usepackage{times}
\usepackage{lineno}
\usepackage{latexsym}
\modulolinenumbers[5]

\begin{document} 

\centerline{\bf Advanced Studies in Theoretical Physics}

\centerline{\bf Vol. 10, 2016, no. 1, 33 - 43}

\centerline{\bf HIKARI Ltd, \ www.m-hikari.com}

\centerline{\bf http://dx.doi.org/10.12988/astp.2016.510103} 

\centerline{} 

\centerline{} 

\centerline{\Large{\bf Constraints on Holographic Cosmological Models}} 

\centerline{} 

\centerline{\Large{\bf from Gamma Ray Bursts}} 

\centerline{} 

\centerline{\bf {Alexander Bonilla Rivera\footnote{e-mail: alex.acidjazz@gmail.com}}}

\centerline{} 

\centerline{Unversidad Distrital Francisco Jos\'e de Caldas}

\centerline{Unversidad Distrital Francisco Jos\'e de Caldas, Carrera 7 No. 40B - 53, Bogot\'a, Colombia.}

\centerline{} 

\centerline{\bf {Jairo Ernesto Castillo Hernandez\footnote{e-mail: jairocastillo63@yahoo.es}}} 

\centerline{} 

\centerline{Unversidad Distrital Francisco Jos\'e de Caldas, Carrera 7 No. 40B - 53, Bogot\'a, Colombia.}

\centerline{}

{\footnotesize Copyright $\copyright$ 2015 Alexander Bonilla Rivera and Jairo Ernesto Castillo Hernandez. This article is distributed under the Creative Commons Attribution License, which permits unrestricted use, distribution, and reproduction in any medium, provided the original work is properly cited.}

\begin{abstract}
\noindent We use Gamma Ray Bursts (GRBs) data from \textit{Y. Wang (2008)} to put additional constraints on a set of cosmological dark energy models based on the holographic principle. GRBs are among the most complex and energetic astrophysical events known in the universe offering us the opportunity to obtain information from the history of cosmic expansion up to about redshift of $z\sim 6$. These astrophysical objects provide us a complementary observational test to determine the nature of dark energy by complementing the information of data from Supernovas (e.g. Union 2.1 compilation). We found that the $\Lambda CDM$ model gives the best fit to the observational data, although our statistical analysis ($\Delta AIC$ and $\Delta BIC$) shows that the models studied in this work (\textit{"Hubble Radius Scale"} and \textit{"Ricci Scale Q"}) have a reasonable agreement with respect to the most successful, except for the \textit{"Ricci Scale CPL"} and \textit{"Future Event Horizon"} models, which can be ruled out by the present study. However, these results reflect the importance of GRBs measurements to provide additional observational constraints to alternative cosmological models, which are mandatory to clarify the way in which the paradigm of dark energy or any alternative model is correct. 
\end{abstract} 


\section{Introduction}

\noindent In order to explain the current acceleration of the universe, the fine-tuning problem of the value of $\Lambda$ and the cosmic coincidence problem, different alternative models have been proposed. In framing the question of the nature of dark energy there are two directions. The firts is to assume a new type of component of energy density, which may be a fluid of constant density or dynamic. The other direction is to modify the Einstein's equations thinking that the metric is inappropriate or that gravity works differently on large scales. The observational tests are of great importance to discriminate between these scenarios \cite{Albrecht:2009ct}. The holographic dark energy is one dynamical DE model proposed in the context of quantum gravity, so called \textit{holographic principle}, which arose from black hole and string theories \cite{2013JCAP...11..053C}. The holographic principle states that the number of degrees of freedom of a physical system, apart from being constrained by an infrared cutoff, should be finite and should scale with its bounding area rather than with its volume. Specifically, it is derived with the help of tha entropy-area relation of thermodynamics of black hole horizons in general relativity, which is also known as the Bekenstein-Hawking entropy bound, i.e., $S\simeq  M^2_p L^2$, where \textit{S} is the maximum entropy of the system of length \textit{L} and $M_p = 1/\sqrt{8 \pi G}$ is the reduced Planck mass. This principle can be applied to the dynamics of the universe, where L may be associated with a cosmological scale and its energy density as:

\begin{equation}
\rho_H = \frac{3c^2 M^2_p}{L^2}.
\end{equation}

\noindent If for example \textit{L} is the Hubble's radius, which represents the current size of the universe, then $\rho_{H}$ represents the holographic dark energy density associated. \\

\noindent Our analysis begin with the holographic dark energy models ($\oint \ref{sec:02}$), following with the Gamma-Ray Bursts model and data ($\oint \ref{sec:03}$) and then finish with the results and discussion ($\oint \ref{sec:04}$).

\section{Holographic Dark Energy Models}\label{sec:02}

\noindent In this section we present some of the most popular holographic dark energy models reported in the literature. \\

The Friedmann equations for a spatially flat universe can be written as:

\begin{equation}
3H =8 \pi G \left( \rho_m + \rho_H  \right),
\end{equation}

\noindent where $\rho_m $ is the energy density of the matter component and $\rho_H$ is the holographic dark energy density. These components are related by an interaction \textit{Q} term as:

\begin{equation}
\frac{d \rho_m}{dt} + 3H\rho_m=Q \qquad \frac{d\rho_{H}}{dt} + 3H(1 + w_H)\rho_{H} = - Q,
\end{equation}

\noindent where $w_H=p_H/\rho_H$ is the equation of state of the holographic dark energy density. The change rate of the Hubble parameter can be written as:

\begin{equation}
\frac{dH}{dt} = - \frac{3}{2} H^{2} \left( 1 + w^{eff} \right), 
\end{equation} 

\noindent where $w^{eff}=w/(1+r)$ is the effective equation of state of the cosmic fluid and $r=\rho_{m}/\rho_{H}$ is the ratio of energy densities, which is related to the saturation parameter $c^2$ as $c^{2}(1 + r)=1$, which establishes that energy in a box of size L should not exceed the energy of the black hole of the same size, under the condition $L^3 \rho_H \leq M^2_p L$. Different scales lead to different cosmological models \cite{2013JCAP...11..053C}.

\subsection{$\Lambda CDM$}\label{sec:02.1}

\noindent We begin our analysis with the standard cosmological model. 
In this paradigm, the DE is provided by the cosmological constant $\Lambda$,  with an EoS, such that $w=-1$. In this model the Friedmman equation $E^{2}(z,\Theta)$ for a flat universe is given by

\begin{equation}
E^{2}(z,\Theta) = \Omega_{r}(1+z)^{4} + \Omega_{m}(1+z)^{3} + \Omega_{\Lambda},
\end{equation} 

\noindent where $\Omega_{m}$ and $\Omega_{\Lambda}$ are the density parameters for matter and dark energy respectively and $\Omega_{r}=\Omega_{\gamma}(1 + 0.2271 N_{eff})$ is the radiation density parameter, where $\Omega_{\gamma}=2.469 \times 10^{-5} h^{-2}$ is photon density parameter and $N_{eff}=3.046$ is the effective number of neutrinos . The free parameters are  \textit{h}, $\Omega_{m}$, $\Omega_{\Lambda}$ and the best fit is shown in Table \ref{tab:LCDM}, 

\begin{table}[htb]
\centering
\begin{tabular}{ll}
\hline 
$\Lambda$ \textit{Cold Dark Matter model} \\ 
\hline 
\hline 
 $h=0.7009\pm 0.0035$ & $\Omega_\Lambda  = 0.716\pm 0.028$ \\ 
$\Omega_m=0.266\pm 0.0042$ &  \\ 
\hline 
\end{tabular} 
\caption{Best fit parameters with all data set to $\Lambda$CDM model.}
\label{tab:LCDM}
\end{table}

\noindent where $h = H_0 /100km.s^{-1}Mpc^{-1}$ is dimensionless Hubble parameter.

\subsection{Hubble Radius Scale}\label{sec:02.2}

\noindent In this model $L=H^{-1}$ and the dark energy density $\rho_H = 3c^2 M^2_p{H^2}$ and the Friedmann equation can be written as:

\begin{equation}
E^{2}(z) =  \left[ (1 - 2q_{0}) + 2(1 + q_{0})(1 + z)^{3n/2} \right]^{1/n} \left( \frac{1}{3}\right)^{1/n},
\end{equation}

\noindent where $q_0$ is the present value of the deceleration parameter. This model is similar to $\Lambda CDM$ when n = 2. The free parameters are \textit{h}, $q_0$ and n, whose best fit values are shown in the Table \ref{tab:HRS}.

\begin{table}[htb]
\centering
\begin{tabular}{ll}
\hline 
 \textit{Hubble Radius Scale} \\ 
\hline 
\hline 
$h=0.7004\pm 0.0038$ &  $n=1.71\pm 0.20$ \\ 
$q_0=0.569 \pm 0.047$ &  \\ 
\hline 
\end{tabular} 
\caption{Best fit parameters with all data set to \textit{HRS} model.}
\label{tab:HRS}
\end{table}

\subsection{Future Event Horizon $\xi =1$}\label{sec:02.3}

\noindent With $L = R_E$ the holographic DE density is $\rho_H = 3c^2 M^2_p{R^{-2}_E}$, where  $R_E$ is the future event horizon. The Friedmann equation is given by:

\begin{equation}
E^{2}(z) =  (1+z)^{3/2-1/c} \sqrt{\frac{1+r_{0}(1+z)}{r_{0}+1}} \left[ \frac{\sqrt{r_{0}(1+z)+1}-1}{\sqrt{r_{0}+1}+1}\right]^{2/c},
\end{equation}

\noindent where $R_E = c \sqrt{(1+r)} H^{-1}$ and $r_0 = \Omega_0 /(1-\Omega_0)$. The best fit values of the free parameters \textit{h}, $r_0$ and \textit{c} are given in the Table \ref{tab:FEH}.

\begin{table}[htb]
\centering
\begin{tabular}{ll}
\hline 
 \textit{Future Event Horizon $\xi =1$} \\ 
\hline 
\hline 
$h=0.6799\pm 0.0025$ &  $r_0=0.322\pm 0.032$ \\ 
$c=1.046\pm 0.017$ &  \\ 
\hline 
\end{tabular} 
\caption{Best fit parameters with all data set to \textit{FEH} model.}
\label{tab:FEH}
\end{table}

\subsection{Ricci Scale CPL}\label{sec:02.4}

\noindent The Ricci scalar $R=6(2H^2 + \dot{H})$ is relate to the cutoff-scale through $L^2 = 6/R$ and the energy:

\begin{equation}
\rho_H = 3c^2 M^2_p \frac{R}{6} = \alpha \left( 2H^2 + \dot{H} \right), 
\end{equation}

\noindent where $\alpha  = 3c^2/8\pi G$.  If we use the CPL parameterization $w(a)=w_0 +(1-a)w_1$, the Friedmann equation can be written as:

\begin{equation}
E^{2}(z,\Theta) =  (1+z)^{\frac{3}{2} \frac{1 + r_{0} + w_{0} + 4 w_{1}}{1 + r_{0} + 3 w_{1}}} \left[ \frac{1 + r_{0} + 3 w_{1} \left( \frac{z}{1 + z}\right) }{1 + r_{0}}\right]^{- \frac{1}{2} \frac{1 + r_{0} + 3 w_{0}}{1 + r_{0} + 3 w_{1}}}.
\end{equation}

\noindent The free parameters of this model are \textit{h}, $r_0$, $w_0$ and $w_1$. The best fit is given in the Table \ref{tab:RSCPL}.

\begin{table}[htb]
\centering
\begin{tabular}{ll}
\hline 
 \textit{Ricci Scale CPL} \\ 
\hline 
\hline 
$h=0.6518\pm 0.0021$ &  $r_0=9.39^{+0.51}_{-0.47} $ \\ 
$w_0=-2.64^{+0.49}_{-0.55} $ & $w_1=0.46^{+0.34}_{-0.33}$  \\ 
\hline 
\end{tabular} 
\caption{Best fit parameters with all data set to \textit{RSCPL} model.}
\label{tab:RSCPL}
\end{table}

\subsection{Ricci Scale Q}\label{sec:02.5}

\noindent If the interaction term is given by $Q=3H\beta \rho_H$, then

\begin{equation}
\beta = \frac{1}{1+r} \left[ rw-\frac{\dot{w}}{H} \right] 
\end{equation}

\noindent and the EoS is given by:

\begin{equation}
w = - \frac{1}{6} \frac{u - s - (u + s) A a^{s}}{1 - A a^{s}}
\end{equation}

\noindent where $u\equiv r_{0}-3w_{0}+3\beta$, $v\equiv r_{0}+3w_{0}+3\beta$, $s\equiv (u^{2}-12\beta(1+r_{0}-3w_{0}))^{1/2}$ y $A\equiv (v-s)/(v+s)$. The Friedmann equation can be writen as:

\begin{equation}
E^{2}(z,\Theta) =   \left[  \frac{n(1 + z)^{-s} - m}{n - m} \right] ^{\frac{3}{2} \frac{lm - kn}{mns}}(1 + z)^{\frac{3}{2} \left( 1 - \frac{k}{m}\right)},
\end{equation}

\noindent  such that $m\equiv 1+r_{0}-1/2(v-s)$, $n\equiv \left[ 1+r_{0}- 1/2(v+s) \right]A$, $k\equiv 1/6(u-s)$ y $l\equiv 1/6 (u+s)A$. The free parameters are \textit{h}, $r_0$, $w_0$ and $\beta$, whose best fit is shown in the Table \ref{tab:RSQ}.

\begin{table}[htb]
\centering
\begin{tabular}{ll}
\hline 
 \textit{Ricci Scale Q} \\ 
\hline 
\hline 
$h=0.6999\pm 0.0038$ &  $r_0=0.201^{+0.025}_{-0.022} $\\ 
$w_0=-0.842^{+0.055}_{-0.056} $ &  $\beta=-0.011^{+0.015}_{-0.019} $  \\ 
\hline 
\end{tabular} 
\caption{Best fit parameters with all data set to \textit{RSCPL} model.}
\label{tab:RSQ}
\end{table}

\section{Gamma-Ray Bursts}\label{sec:03}

In this section we present a brief introduction of GRBs as astrophysical objects and its later use in cosmology.

\subsection{GRBs Model}\label{sec:03.1}

\noindent Gamma-ray bursts (GRBs) are the most luminous astrophysical events observable today, because they are at cosmological distances. The duration of a gamma-ray burst is typically a few seconds, but can range from a few milliseconds to several minutes. The initial burst at gammay-ray wavelengths is usually followed by a longer lived afterglow at longer wavelengths (X-ray, ultraviolet, optical, infrared, and radio). Gamma-ray bursts have been detected by orbiting satellites about two to three times per week. Most observed GRBs appear to be collimated emissions caused by the collapse of the core of a rapidly rotating, high-mass star into a black hole. At least once a day, a powerful source of gamma rays temporarily appears into the sky in an unpredictable location and later disappears, which last for milliseconds to minutes. In the location of the gamma ray event is usually observed a dominant afterglow in X-rays, optical and radio after long decays. 

\subsection{GRBs Data}\label{sec:03.2}

\noindent We use GRB data in the form of the model-independent distance from Wang (2008) \cite{2008PhRvD..78l3532W}, which were derived from the data of 69 GRBs with $0.17 \leq z \leq 6.6$ from Schaefer (2007). The GRB data are included in our analysis by adding the following term to the given model:

\begin{equation}
\chi^2_{\scriptscriptstyle GRB} =
\left[ \Delta \bar{r}_{p}(z_i) \right] . (C_{\scriptscriptstyle
GRB}^{-1})_{ij}.\left[ \Delta \bar{r}_{p}(z_i) \right], 
\end{equation}

\noindent where $\Delta \bar{r}_{p}(z_i)=\bar{r}_{p}^{data}(z_i)-\bar{r}_{p}(z_i)$ and $\bar{r}_{p}(z_i)$ is given by

\begin{equation}
\bar{r}_{p}(z_i)=\frac{r_{p}(z)}{r_{p}(0.17)}
\end{equation}

\noindent where

\begin{equation}
r_{p}(z)=\frac{(1+z)^{1/2}}{z} \frac{H_0}{ch}r(z)
\end{equation}

\noindent and $r(z)$ is the comoving distance at \textit{z}. The covariance matrix is given by:

\begin{equation}
C^{\scriptscriptstyle
grb}_{ij} = \sigma (\bar{r}_{p}(z_i)) \sigma (\bar{r}_{p}(z_j)) \bar{C}^{\scriptscriptstyle
grb}_{ij}
\end{equation}

\noindent where $\bar{C}^{\scriptscriptstyle grb}_{ij}$ is the normalized covariance matrix:\vspace{4pt}\\

\begin{equation}
\bar{C}^{\scriptscriptstyle grb}_{ij} = \left(
\begin{array}{cccccc}
1.0000 &              &             &              &              &            \\
0.7056 & 1.0000 &              &              &             &             \\
0.7965 & 0.5653 & 1.0000 &              &              &             \\
0.6928 & 0.6449 & 0.5521 & 1.0000 &              &             \\
0.5941 & 0.4601 & 0.5526 & 0.4271 & 1.0000 &              \\
0.5169 & 0.4376 & 0.4153 & 0.4242 & 0.2999 & 1.0000 \\
\end{array}\right)
\end{equation}

\noindent and

\begin{equation}
 \sigma (\bar{r}_{p}) = \left\lbrace 
 \begin{array}{l}
\sigma (\bar{r}_{p}(z_i))^+, \hspace{0.5cm}  if \hspace{0.5cm} \bar{r}_{p}(z_i)\geq \bar{r}_{p}(z_i)^{dat}\medskip \\ 
\sigma (\bar{r}_{p}(z_i))^-,\hspace{0.5cm}   if  \hspace{0.5cm}  \bar{r}_{p}(z_i) < \bar{r}_{p}(z_i)^{dat}
\end{array}
\right. 
\end{equation}

\noindent where $\sigma (\bar{r}_{p}(z_i))^+$ and $\sigma (\bar{r}_{p}(z_i))^-$, are $68\%$ C.L errors given in the Table \ref{tab:GRBSig}. As complementary tests we use SNIa (580-Data point), CMB (1-Data point) and BAO (1-Data point)  \cite{2007JCAP...01..018N} (See Appendix).

\begin{table}
\centering
\begin{tabular}{c|cccc}
\hline
\hline
Data point & (\textit{z}) & $\bar{r}_{p}(z_i)^{dat}$ & $\sigma (\bar{r}_{p}(z_i))^+$ & $\sigma (\bar{r}_{p}(z_i))^-$\\
\hline
0 & 0.17   & 1.0000 & -           &  -          \\

1 & 1.036 & 0.9416 & 0.1688 & 0.1710 \\

2 & 1.902 & 1.0011 & 0.1395 & 0.1409 \\

3 & 2.768 & 0.9604 & 0.1801 & 0.1785 \\

4 & 3.634 & 1.0598 & 0.1907 & 0.1882 \\

5 & 4.500 & 1.0163 & 0.2555 & 0.2559 \\

6 & 6.600 & 1.0862 & 0.3339 & 0.3434 \\
\hline
\end{tabular}
\caption{Distances independent model GRBs.}
\label{tab:GRBSig}
\end{table}

\section{Results and Discussion}\label{sec:04}

In this section we perform the statistical analysis, where we implemented the maximum likelihood criterion to get the best settings for each model and used the AIC and BIC criterion to discriminate between the different models.\\

\noindent The maximum likelihood estimate for the best fit parameters  is:

\begin{equation}
 \mathcal{L}_{max}= exp \left[ -\frac{1}{2}{\chi_{min}^{2}} \right] 
\end{equation}

\noindent If $\mathcal{L}_{max}$ has a Gaussian errors distribution, then $\chi_{min}^{2}=-2 \ln \mathcal{L}_{max}$, So, for our analysis:

\begin{equation}
\chi_{min}^{2}= \chi^{2}_{GRBs}+\chi^{2}_{SNIa}+\chi^{2}_{CMB} +\chi^{2}_{BAO}.
\end{equation}

\begin{figure}[htb]
   \centering
   {\fboxrule=0.5pt \fboxsep=5pt
    \fbox{
    \includegraphics[width=0.45\textwidth,angle=0]{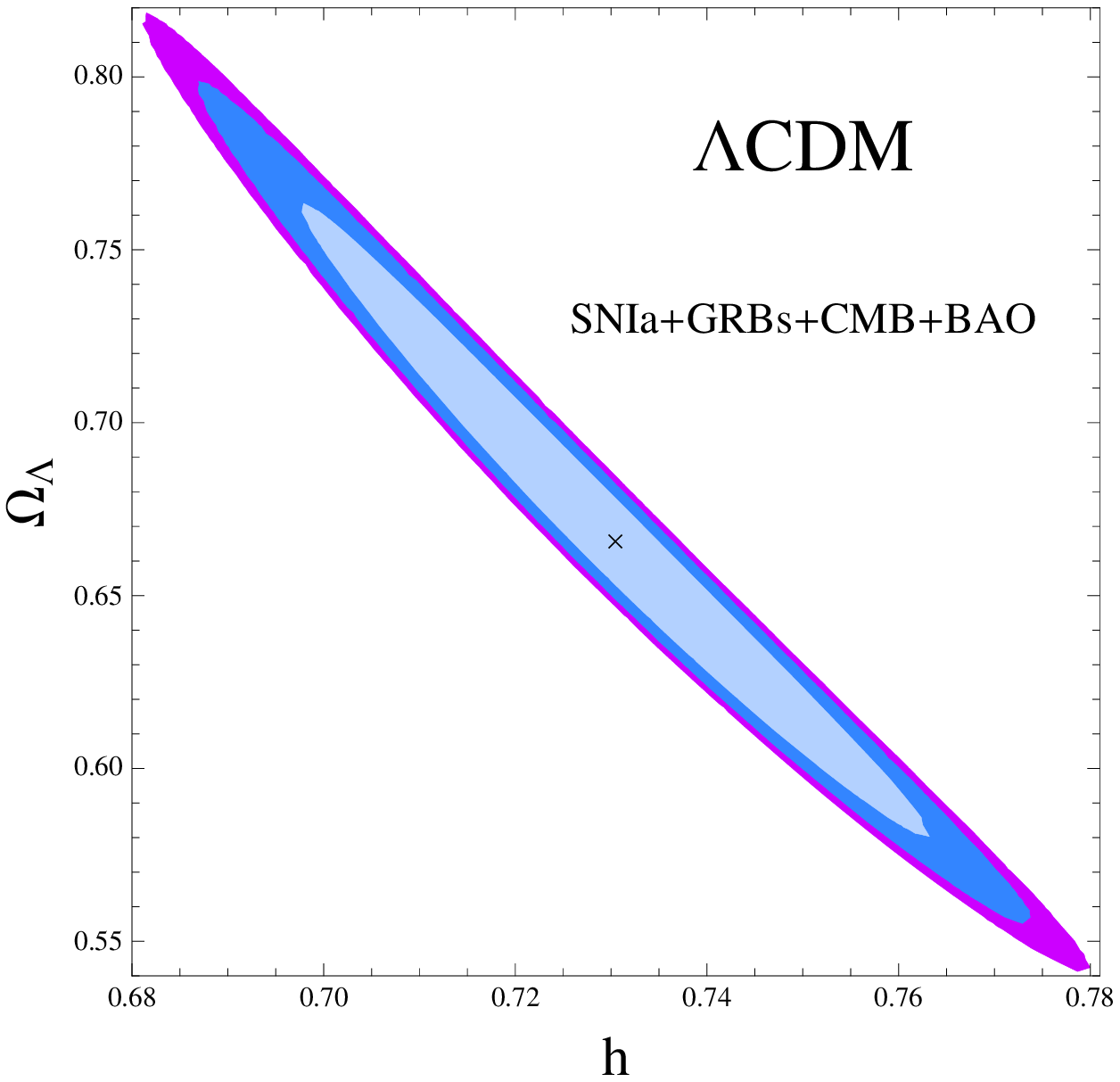}
     \includegraphics[width=0.45\textwidth,angle=0]{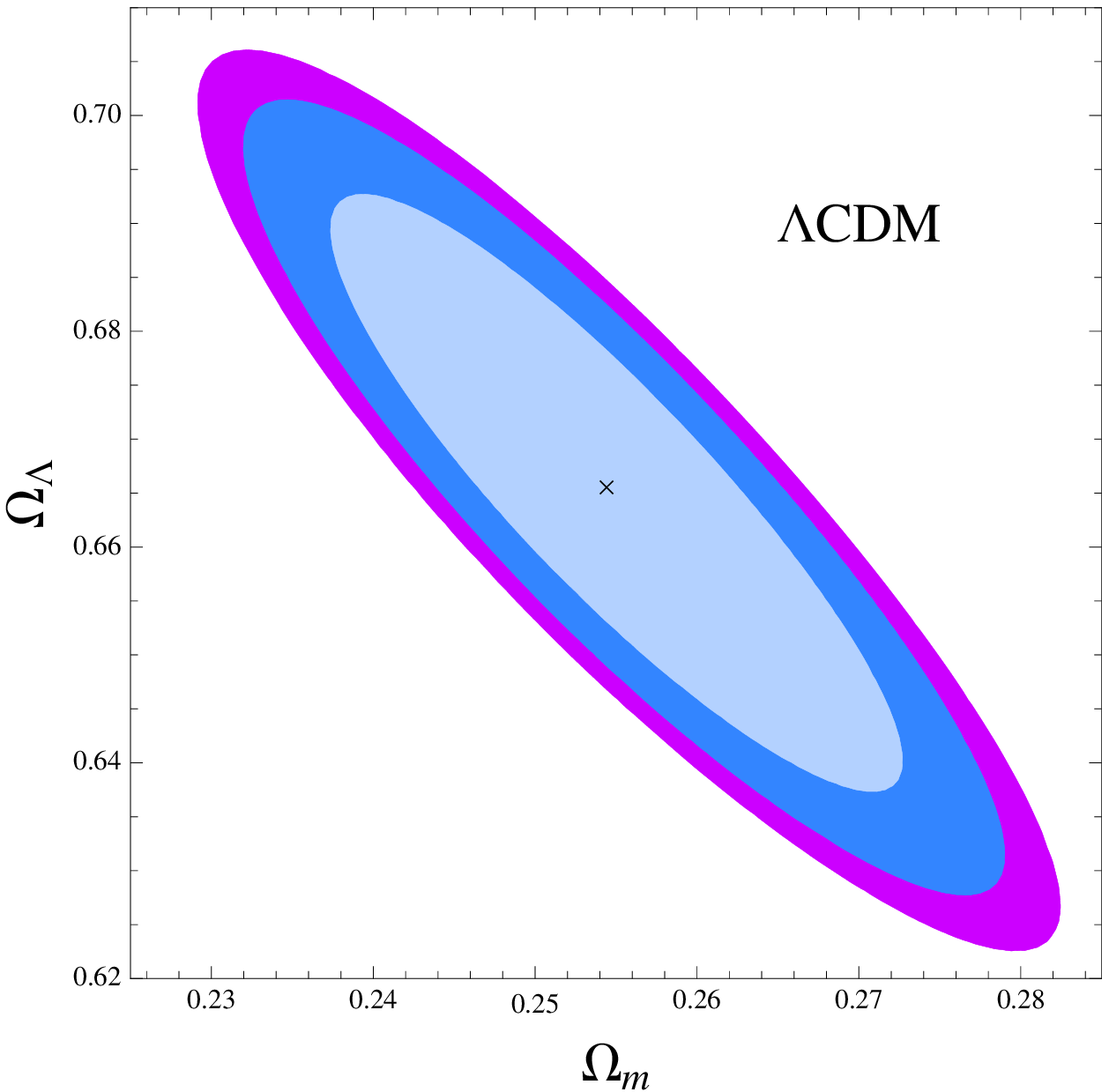}
     }}
     {\fboxrule=0.5pt \fboxsep=5pt
    \fbox{
     \includegraphics[width=0.45\textwidth,angle=0]{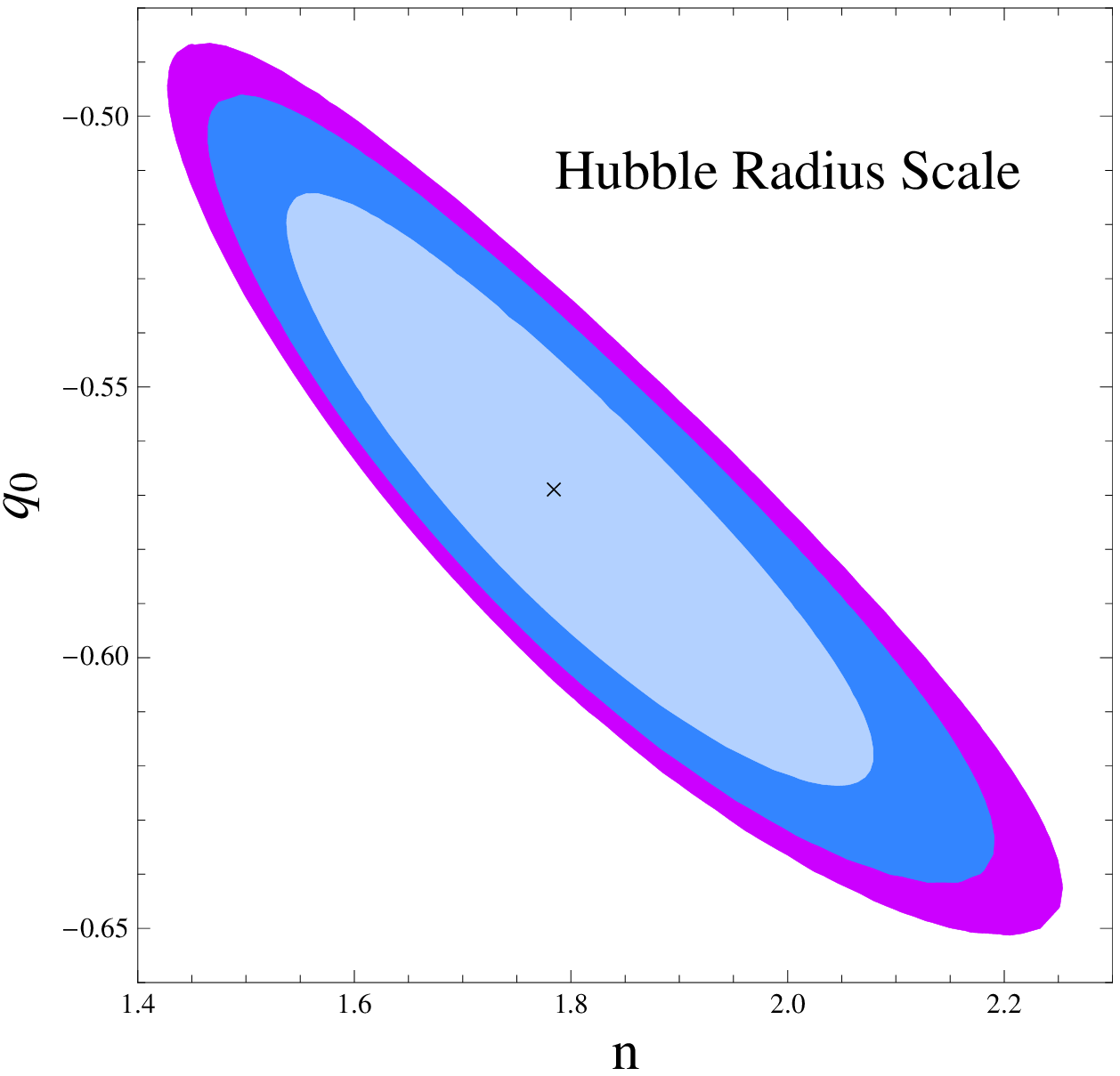}
     \includegraphics[width=0.45\textwidth,angle=0]{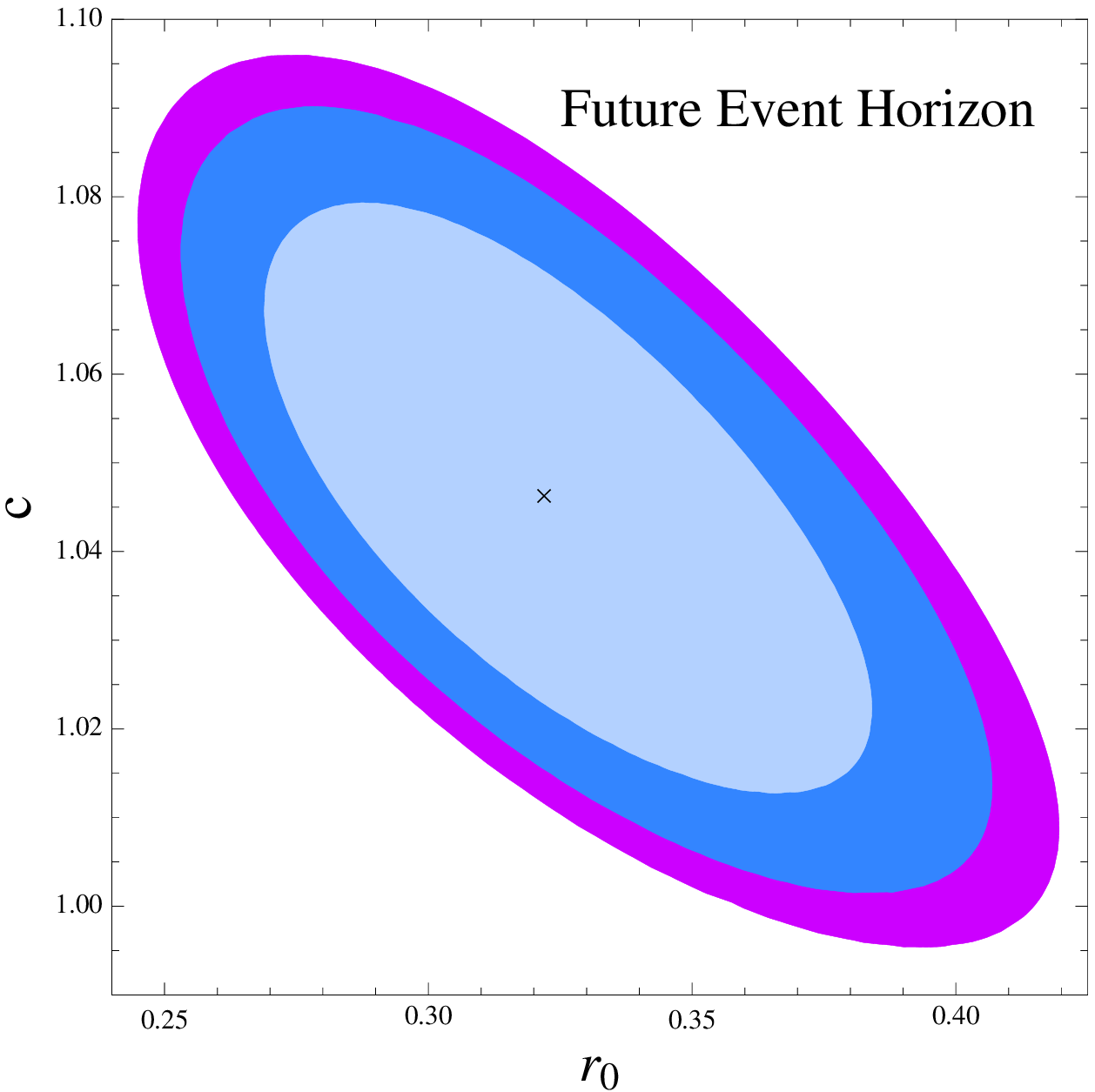}
     }}
       {\fboxrule=0.5pt \fboxsep=5pt
    \fbox{
     \includegraphics[width=0.45\textwidth,angle=0]{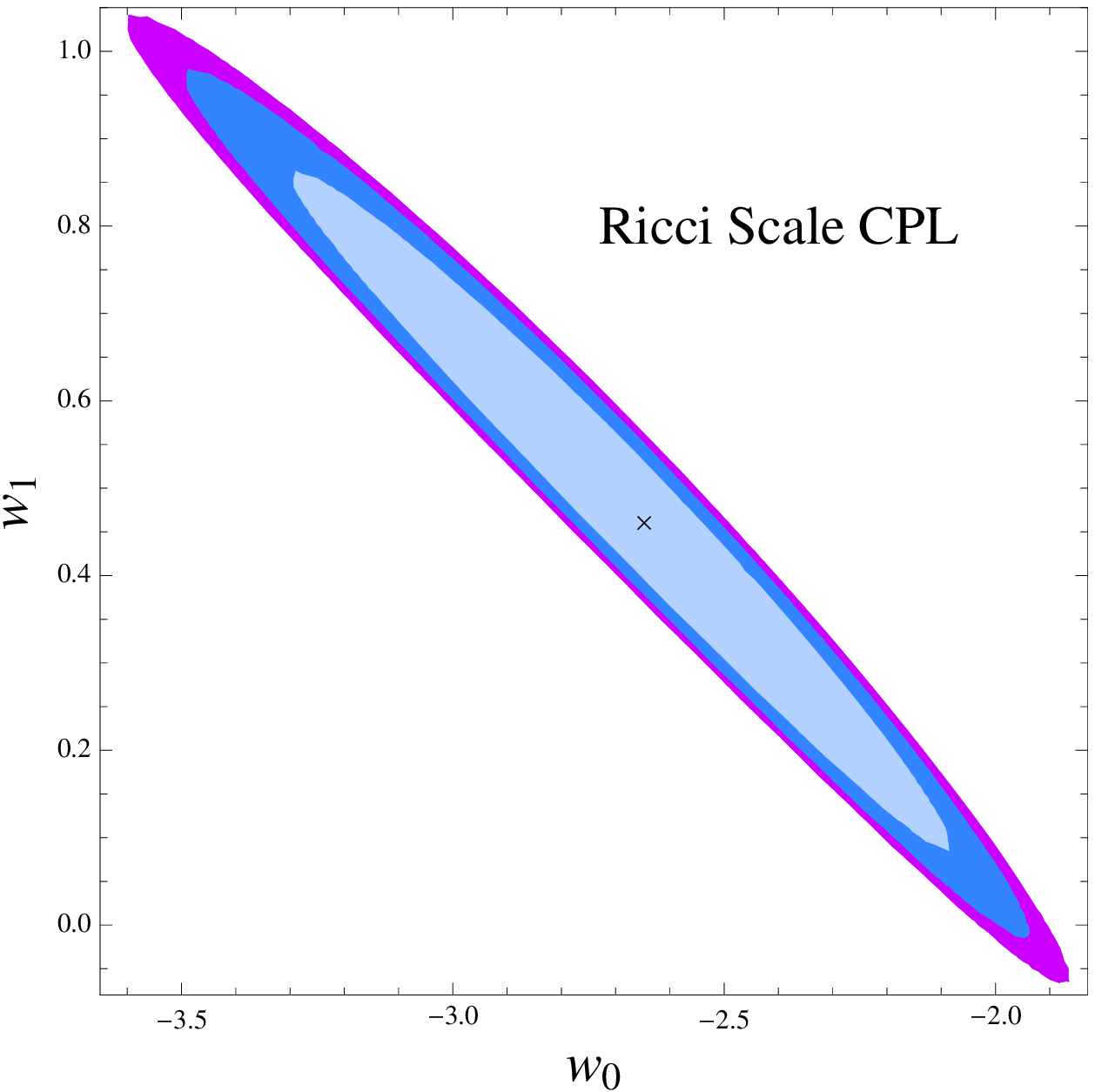}
     \includegraphics[width=0.45\textwidth,angle=0]{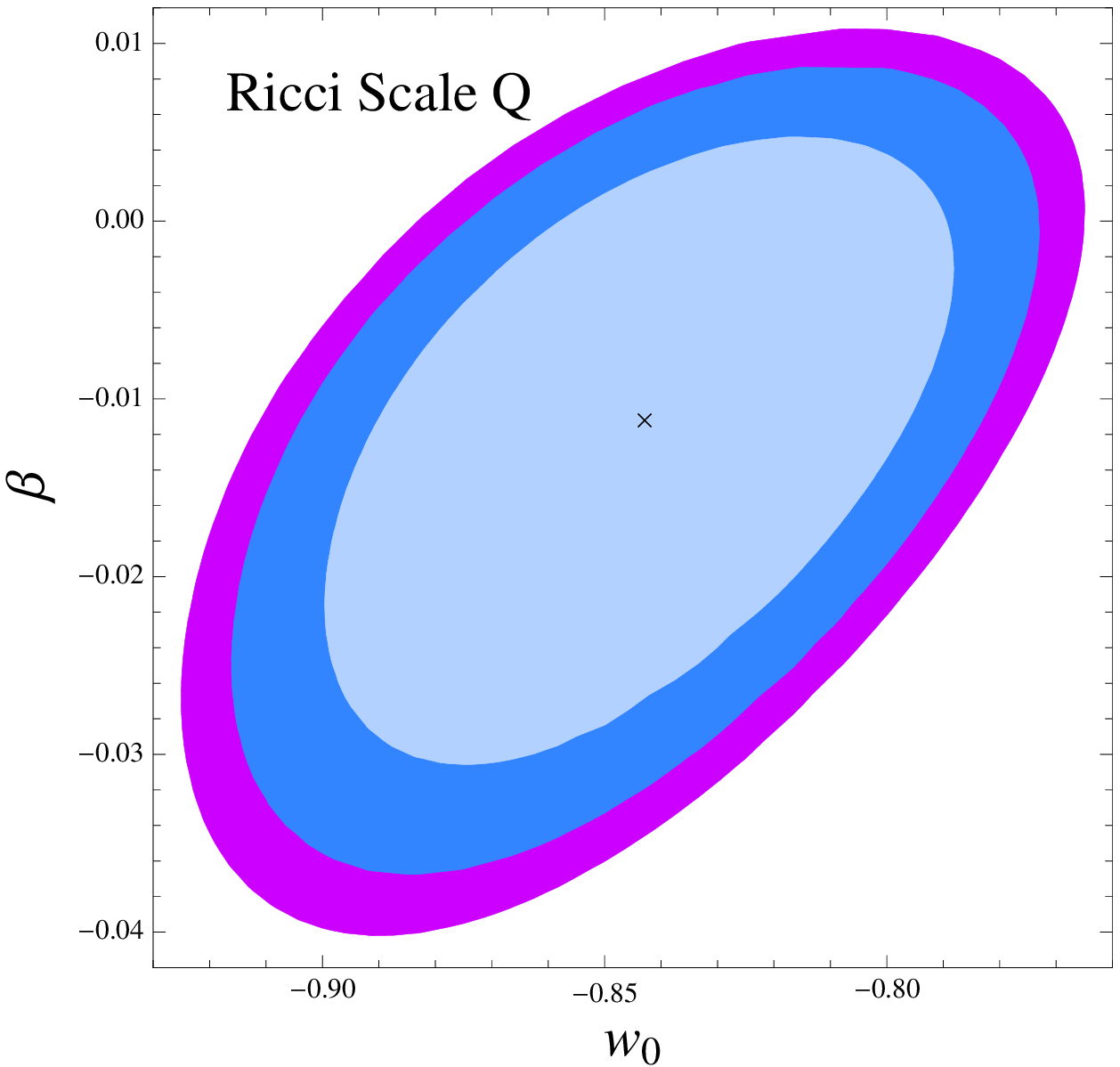}
     }}
    \caption{Diagrams of statistical confidence marginalizing different cosmological parameters at $1\sigma$, $2\sigma$ and $3\sigma$ for different cosmological models.}\label{figure:ConDia}
\end{figure}

\noindent In the Figure \ref{figure:ConDia} we  show the diagrams of statistical confidence at $1\sigma$, $2\sigma$ and $3\sigma$ for different cosmological models and several parameter space, from a joint analysis of 69 GRBs (independent-model 6-Data point),  SNIa (580-Data point), CMB (1-Data point), BAO (1-Data point) \cite{2007JCAP...01..018N} \cite{2013JCAP...11..053C}.\\ 

\noindent In this paper we use the Akaike and Bayesian information criterion 
(AIC, BIC), which allow to compare cosmological models with different degrees of freedom, 
with respect to the observational evidence and the set of parameters  \cite{Schwarz:78}. The AIC and BIC can be calculated as:

\begin{equation}
AIC=-2 \ln \mathcal{L}_{max} +2k,
\end{equation}

\begin{equation}
BIC=-2 \ln \mathcal{L}_{max} + k \ln N, 
\end{equation}

\noindent where $\mathcal{L}_{max}$ is the maximum likelihood of the model under consideration, 
$k$ is the number of parameters.  BIC imposes a strict penalty against extra parameters for any set with \textit{N} data. The prefered model is that which minimizes AIC and BIC, however,  only the relative values between the different models is important \cite{Liddle:04}. The results are shown in the Table \ref{tab:AICBIC}.

\begin{table}{}
\centering
\begin{tabular}{lccccc}
\hline 
\hline 
       Model                         & $\chi^2_{min}$ & \textit{AIC} & \textit{BIC}& $\Delta AIC$ & $\Delta BIC$ \\ 
\hline 
$\Lambda CDM$               & 608.5 & 614.5 & 627.6  & 0.0 & 0.0\\ 

\textit{Hubble Radius S.}   & 609.7 & 615.7 & 628.8 & 1.2 & 1.2 \\ 

\textit{Future Event H.}    & 657.3 & 663.3 & 676.4 & 48.8 & 48.8 \\ 

\textit{Ricci scale CPL}      & 917.3 & 925.3 & 924.8 & 310.8 & 315.2 \\ 

\textit{Ricci scale Q}          & 609.8 & 617.8 & 635.3 & 3.3 & 7.7 \\ 
\hline 
\end{tabular}
\caption{AIC and BIC analysis to diferent dark energy models using all data sets.}
\label{tab:AICBIC}
\end{table}

\section{Conclusion} \label{sec:05}
\noindent We implement GRBs data model-independent from \textit{Y. Wang (2008)} to complement SNIa Union2.1 sample to high redshift. We found that model-independent GRBs provide additional observational constraints to  holographic dark energy models. In addition to the data of GRBs we use data from SNIa, BAO and CMB as usual cosmological tests.\\

\noindent Our analysis shows that the $\Lambda CDM$ model is preferred by the \textbf{$\Delta AIC$} and \textbf{$\Delta BIC$} criterion. The \textit{"Hubble Radius Scale"} and \textit{"Ricci Scale Q"} models show an interesting agreement with the observational data. The \textit{"Ricci Scale CPL"} and \textit{"Future Event Horizon"} models can be ruled out by the present analysis. \\

\noindent Finally we want to highlight the importance of deepening in the development of unified models of GRBs, given the obvious importance of these objects in the era of cosmology accuracy, which help to shed light on the dark energy paradigm.\\


{\bf Acknowledgements.} A. Bonilla and J. Castillo wish to express their gratitude to Universidad Distrital FJDC  for the academic support and funding. A. Bonilla wishes to thank to professor \textit{Arthur Kosowsky (University of Pittsburgh, U.S.A.)\footnote{{http://www.physicsandastronomy.pitt.edu/people/arthur-kosowsky}}} for his comments and suggestions to this work within the framework of \textit{Centenary Celebration of General Relativity Theory: Andean School on Gravity and Cosmology} \footnote{{http://gravityschool.uniandes.edu.co/index.php/en/speakers}}. A. Bonilla also wish give thanks to \textit{Dr. Ga\"{e}l Fo\"{e}x \footnote{{http://ifa.uv.cl/index.php/es/2-uncategorised/83-dr-gael-foeex}}} for their appropriate comments and review of this work.


\def \aap {A\&A} 
\def \aapr {A\&AR} 
\def \statisci {Statis. Sci.} 
\def \physrep {Phys. Rep.} 
\def \pre {Phys.\ Rev.\ E.} 
\def \sjos {Scand. J. Statis.} 
\def \jrssb {J. Roy. Statist. Soc. B} 
\def \pan {Phys. Atom. Nucl.} 
\def \epja {Eur. Phys. J. A} 
\def \epjc {Eur. Phys. J. C} 
\def \jcap {J. Cosmology Astropart. Phys.} 
\def \ijmpd {Int.\ J.\ Mod.\ Phys.\ D} 
\def \nar {New Astron. Rev.} 

\def \JCAP {JCAP}
\def \araa {ARA\&A}
\def \aj {AJ}
\def \aar {A\&AR}
\def \apj {ApJ}
\def \apjl {ApJL}
\def \apjs {ApJS}
\def \asl {Adv. Sci. Lett.} 
\def \mnras {Mon.\ Non.\ Roy.\ Astron.\ Soc.}
\def \nat {Nat}
\def \pasj {PASJ}
\def \pasp {PASP}
\def \science {Science}

\def \gca {Geochim.\ Cosmochim.\ Acta}
\def \npa {Nucl.\ Phys.\ A}
\def \plb {Phys.\ Lett.\ B}
\def \prc {Phys.\ Rev.\ C}
\def \prd {Phys.\ Rev.\ D.}
\def \prl {Phys.\ Rev.\ Lett.}


\section*{Appendix}\label{sec:06}


\subsection*{SNIa}\label{sec:06.1}
Here, we use the Union $2.1$ sample which contains 580 data points.
The SNIa data give the luminosity distance $d_L(z)=(1+z)r(z)$. We
fit the SNIa with the cosmological model by minimizing the $\chi^2$
value defined by

\begin{equation}
\chi_{SNIa}^2=\sum_{i=1}^{580}\frac{ [\mu(z_{i})-\mu_{obs}(z_{i})]^2 }{\sigma_{\mu_i}^2},
\end{equation}

 where  $\mu(z)\equiv 5\log_{10}[d_L(z)/\texttt{Mpc}]+25$ is the
theoretical value of the distance modulus, and $\mu_{obs}$ the observed one.

\subsection*{CMB}

We also include CMB information by using the WMAP  data. The $\chi^2_{cmb}$ for the CMB data is constructed as:

\begin{equation}
\chi^2_{cmb} = \frac{\left(  1.7246 - R \right)^2 }{0.03^2}. 
\end{equation}

Here $R$ is ``shift parameter'',  defined as:
\begin{equation}
R = \frac{\sqrt{\Omega_{m}}}{c(1+z_*)} D_L(z).
\end{equation}

where $d_L(z)=D_L(z)/H_0$ and the redshift of decoupling $z_*$ is
 $z_* = 1048[1+0.00124(\Omega_b h^2)^{-0.738}] [1+g_1(\Omega_{m}h^2)^{g_2}]$ and

\begin{equation}
g_1 = \frac{0.0783(\Omega_b h^2)^{-0.238}}{1+39.5(\Omega_b
h^2)^{0.763}},
 g_2 = \frac{0.560}{1+21.1(\Omega_b h^2)^{1.81}}.
\end{equation}

\subsection*{BAO}

Similarly, for the DR7 BAO data, the $\chi^2$ can be expressed as:

\begin{equation}
\chi^2_{\scriptscriptstyle 6dFGS} =
\left(\frac{d_z-0.469}{0.017}\right)^2.
\end{equation}

where $d_z = r_s(z_d)/D_V(z)$ denotes the distance ratio. Here, $r_s(z_d)$ is the comoving sound horizon at the baryon drag epoch ($z_d=0.35$) and $D_V(z)$ is the acoustic scale.

\centerline{}

{\bf Received: November 6, 2015; Published: January 21, 2016}


\begin{thebibliography}{00} 


\bibitem[1]{Albrecht:2009ct} 
 A. Albrecht, L. Amendola, G. Bernstein, D. Clowe, D. Eisenstein, L. Guzzo, C. Hirata and D. Huterer {et al.}, {Findings of the Joint Dark Energy Mission Figure of Merit Science Working Group}, 2009, [arXiv:0901.0721].

\bibitem[2]{2013JCAP...11..053C} V. H. C{\'a}rdenas, A. Bonilla, V. Motta, S. del Campo, Constraints on holographic cosmologies from strong lensing systems, {\it \jcap}, (2013), no. 11, 053. http://dx.doi.org/10.1088/1475-7516/2013/11/053 

\bibitem[3]{Liddle:04} A. R. Liddle, {How many cosmological parameters?}, 2004, MNRAS 351, L49-L53, [astro-ph/0401198v3].

\bibitem[4]{2007JCAP...01..018N} S. Nesseris, L. Perivolaropoulos, Crossing the phantom divide: theoretical implications and observational status, {\it \jcap}, (2007), no. 1, 018. http://dx.doi.org/10.1088/1475-7516/2007/01/018 

\bibitem[5]{Schwarz:78}  G. Schwarz, {Estimating the Dimension of a Model},  \textit{The Annals of Statistics},  6 (1978), 461-464. http://dx.doi.org/10.1214/aos/1176344136 

\bibitem[6]{2008PhRvD..78l3532W} Y. Wang, Model-independent distance measurements from gamma-ray bursts and constraints on dark energy, {\it \prd}, 78 (2008), 123532. http://dx.doi.org/10.1103/physrevd.78.123532

\end{thebibliography}
\end{document}